# Freezing as a Path to Build Complex Composites

Sylvain Deville, Eduardo Saiz, Ravi K. Nalla**, and Antoni P. Tomsia*

*Materials Sciences Division, Lawrence Berkeley National Laboratory, Berkeley, CA 94720, USA*

*** now with Intel Corporation, 5000 W. Chandler Blvd. CH5-159, Chandler, AZ-85226, USA*

**One-Sentence Summary:** Freezing is harnessed to create composites that replicate the intricate structure of nacre and to synthesize porous bone substitutes with high strength.

**Abstract:** Materials that are strong, ultra-light weight and tough are in demand for a range of applications, requiring architectures and components carefully designed from the micrometer down to nanometer scales. Nacre—a structure found in many molluscan shells—and bone are frequently used as examples for how nature achieves this through hybrid organic-inorganic composites. Unfortunately, it has proven extremely difficult to transcribe nacre-like clever designs into synthetic materials, partly because their intricate structures need to be replicated at several length scales. We demonstrate how the physics of ice formation can be used to develop sophisticated porous and layered-hybrid materials, including artificial bone, ceramic/metal composites, and porous scaffolds for osseous tissue regeneration with strengths up to four times higher than those currently used for implantation.

---

* Corresponding author: Tel: (510) 486-4918; Fax: (510) 486-4761; E-mail address: APTomsia@lbl.gov

**Text**:

While the potential of layered materials has long been recognized (1), their creation requires solving a two-fold problem, namely the design of optimum microstructures, and development of fabrication procedures to implement these designs. Natural materials such as nacre offer a wealth of information to guide such a design process (2, 3). The unique properties of natural layered materials are achieved through a fine control of the layer thickness, selection of the right components, and manipulation of roughness and adhesion at the organic-inorganic interface (4, 5). The ideal fabrication process has to be not only simple, but also adaptable enough to fabricate layers with a large number of material combinations and a wide range of layer dimensions. Previous techniques for mimicking nacre are bottom-up chemical approaches (6, 7) that are intrinsically limited to a narrow range of materials exhibiting the proper interfacial reactions and compatibility. Other techniques offer only a coarse control of the layer thickness or have practical limitations regarding the number of layers that can be fabricated (7, 8).

In sea ice, pure hexagonal ice platelets with randomly oriented horizontal *c* crystallographic axes are formed, and the various impurities originally present in sea water (salt, biological organisms, etc.) are expelled from the forming ice and entrapped within channels between the ice crystals (9). We apply this natural principle to ceramic particles dispersed in water to build sophisticated, nacre-like architectures in a simple two-step approach. Ice-templated (IT), porous, layered materials—with layers as small as 1 μm—are first fabricated through a freeze-casting process, which involves the controlled unidirectional freezing of ceramic suspensions. These porous scaffolds are then filled with a selected second phase (organic or inorganic) to fabricate dense composites. Using a natural, self-organizing phenomenon, we allow nature to guide the design and processing.

The physics of water freezing has drawn the attention of scientists for a long time. With few exceptions (10), much of this work has concentrated on the freezing of pure water or very dilute suspensions (9, 11). This phenomenon is critical for various applications, such as cryo-preservation of biological cell suspensions (12) and the purification of pollutants (13). An important observation in these studies is that, during the freezing of such suspensions, there is a critical particle size (11) above which the suspended particles will be trapped by the moving water-ice front. Another important observation is that the hexagonal ice crystals exhibit strong anisotropic growth kinetics, varying over about two orders of magnitude with crystallographic orientation. Under steady-state conditions, it is possible to grow ice crystals in the form of platelets, with a very high aspect ratio. The ice thus formed will have a lamellar microstructure, with the lamellae thickness depending mainly on the speed of the freezing front. We designed a simple experimental setup (Fig. S1) that allowed us to precisely control the freezing kinetics. By freezing concentrated suspensions containing ceramic particles with suitable granulometry, we are able to build homogeneous, layered, porous scaffolds.

In the method proposed here (supporting online text), directional freezing of the ceramic slurries is achieved by pouring them into Teflon molds placed between two copper cold fingers (Fig. S1) whose temperature is regulated to control the speed of the solidification-front (Fig. S2). As in nature, during the freezing of sea water, the ceramic particles concentrate in the space between the ice crystals (Fig. 1A). When the freezing rate increases, the magnitude of supercooling ahead of the solidifying interface is increased and, as a result, the tip radius of the ice crystals decreases. A finer microstructure is thus obtained without affecting the long-range order of the entire structure. Afterwards, the ice is sublimated by freeze drying, such that a ceramic scaffold whose microstructure is a negative replica of the ice is produced (Fig. 1F). We controlled the growth of lamellar ice by adjusting the freezing kinetics. In this way, we achieved a layered microstructure with relevant dimensions that vary over two decades (Fig. 2A), from 1 μm (almost the same as nacre, typically ~0.5 μm) (14) to 200 μm, without affecting the ordered architecture. To a large extent, the mesostructure of natural materials determines their mechanical response (15), and it has been difficult to replicate this synthetically. Our results indicate that, by controlling the freezing kinetics and patterns of the cold finger, it is also possible to build mesostructural features and gradients (Fig. 2F) that could optimize the mechanical response of the final materials—for example, by stiffening the structure and limiting torsion (15).

The IT porous scaffolds obtained by this process exhibit striking similarities to the meso- and micro-structure of the inorganic component of nacre (Fig. 1). The inorganic layers are parallel to each other and very homogeneous throughout the entire sample (Fig. 1E). Particles trapped in between the ice dendrites (Fig. 1A) lead to a dendritic surface roughness of the walls (Fig. 1F), just as in nacre (Fig. 1D)(16). Finally, some dendrites span the channels between the lamellae (Fig. 2B), mimicking the tiny inorganic bridges linking the inorganic platelets of nacre, which are believed to increase the fracture resistance (17).

The inorganic portion represents 95% of the volume of nacre, but its highly specific properties (in particular its great toughness) are due to the interaction of this inorganic component with the organic phase (protein) that is found between the calcium carbonate platelets (14). To obtain similar synthetic materials, the next step is to fill the porous ceramic scaffolds with a second phase. For example, we filled the IT scaffolds with a simple organic phase (epoxy) or with an inorganic component (metal, Fig. 1E). Nature shows that the optimum fracture properties are encountered not only when the organic/inorganic interface is strong, but also when delamination at the organic/inorganic interface occurs before the crack goes across the stiff, brittle layer. In the IT composites, extensive crack deflection at the interface between layers was observed (Figs 3B and 3D). As in nacre (Fig 3C), this delamination creates tortuous cracks that propagate in a stable manner (Fig 3A) and increases the toughness of the materials. It is believed that nature manipulates adhesion in two ways—mechanical and chemical. In nacre, this is done by controlling the roughness and the highly specific properties of the polymer adhesive phase (4). Our process allows us to control the morphology of the inorganic layers and the chemistry of the interface. For example, the mechanical response of alumina/Al-Si (45/55 vol.%) layered composites (Fig. 1E) can be manipulated by controlling the interfacial bonding. By adding as little as 0.5 wt.% Ti (known to segregate at the metal/ceramic interfaces (18)) to the aluminum eutectic, the strength increases from 400 to 600 MPa and fracture toughness from 5.5 to 10 MPa$\sqrt{m}$ (Fig. 3E).

Our technique shows promise for a large number of applications that require tailored composite materials. One such scientific challenge that could be solved is the development of new biomaterials for orthopaedic applications (19). Despite extensive efforts in the development of porous scaffolds for bone regeneration, all porous materials have an inherent lack of strength associated with porosity. By applying freezing to commercial hydroxyapatite (HAP, the mineral component of bone) powder suspensions, we processed IT highly porous lamellar scaffolds that are four times stronger in compression than conventional porous HAP (Fig. 4). These IT scaffolds exhibit well-defined pore connectivity along with directional and completely open porosity of an adequate size to allow bone ingrowth (20). Hence, most of the current shortcomings (low strength, random organization, multiple pore size, uncontrolled pore connectivity) that plague such bone substitutes are eliminated by this innovative approach.

Current ceramic and metallic implant materials have serious shortcomings due to the mismatch of physical properties with those of bone. In bone, intrinsically weak materials, such as calcium phosphates and collagen, are combined into composites exhibiting intermediate modulus (10-20 GPa), fairly high strength (30 to 200 MPa), and high work of fracture (100-1000 J/m$^2$) (21). The unique properties of bone arise from the controlled integration of the organic (collagen) and inorganic (apatite) components (5) with a sophisticated architecture from the nano- to meso-levels. Our approach to the problem is to infiltrate the IT porous HAP scaffolds with a second organic phase with tailored biodegradability. Because the biodegradation rates of the scaffold and the infiltrated compound can be designed to be different, porosity will be created in situ to allow bone ingrowth. Using this approach, we have been able to fabricate HAP-based composites, with stiffness (10 GPa), strength (150 MPa), and work of fracture (220 J/m$^2$) that match that of compact bone for an equivalent mineral/organic content (around 60/40 vol.%).

**Acknowledgments**

This work was supported by the National Institute of Health (NIH/NIDCR) under grant No. 5R01 DE015633 (Novel Scaffolds for Tissue Engineering and Bone-Like Composites) and by the Director, Office of Science, Office of Basic Energy Sciences, Division of Materials Sciences and Engineering of the Department of Energy under Contract No. DE-AC03-76SF00098 (Metal/Ceramic Composites). The authors wish to thank Prof. Robert O. Ritchie for useful discussions and James Wu for help with the synthesis of the aluminum-infiltrated composites.

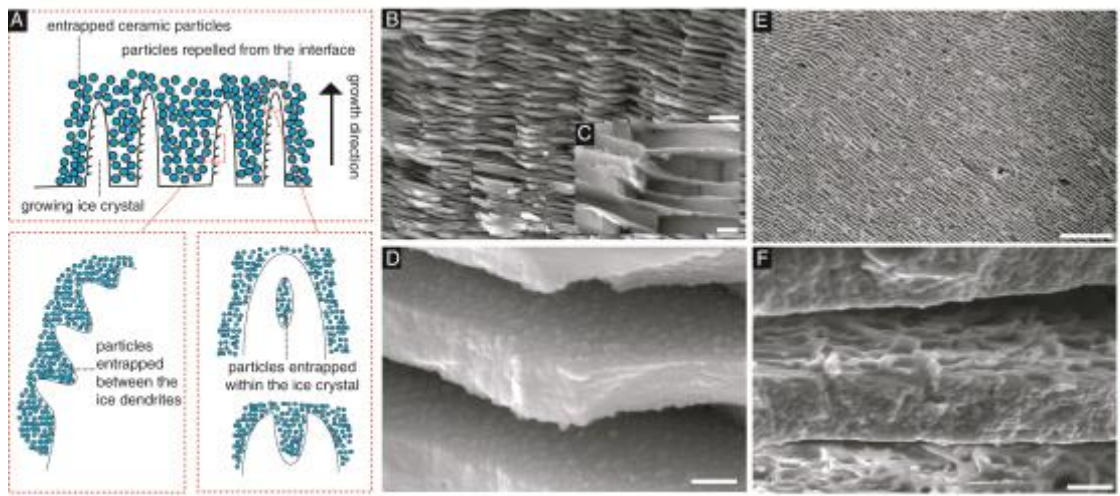

**Figure 1**. **Processing principles and materials**. While the ceramic slurry is freezing, the growing ice crystals expel the ceramic particles, creating a lamellar microstructure oriented in a direction parallel to the movement of the freezing front (A). For highly concentrated slurries, the interaction between particles becomes critical: a small fraction of particles are entrapped within the ice crystals by tip-splitting and subsequent healing (A), leading to the formation of inorganic bridges between adjacent walls. Dense composites are obtained by infiltrating the porous lamellar ceramic with a second phase (e.g., a polymer or a liquid metal). Natural nacre has a brick-mortar-bridges microstructure where inorganic calcium carbonate layers are held together by organic protein "glue" (B and C), the roughness of the inorganic walls (D) is a key contributor to the final mechanical properties of nacre. The layered microstructure of the IT dense composites resembles that of nacre (for example the alumina/Al-Si composite in E). The particles entrapped between the ice dendrites generate a characteristic roughness on the lamella surface (F) that mimics that of the inorganic component of nacre. Scale bars (B) 5 µm (C) 0.5 µm (D) 0.3 µm (E) 300 µm (F) 10 µm.

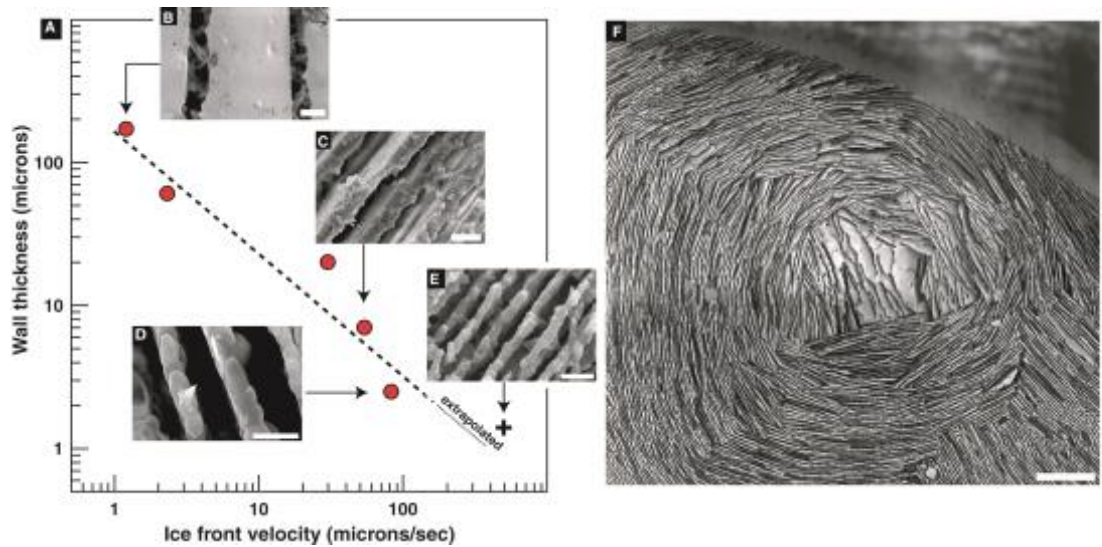

**Figure 2**. M**icrostructural control at several levels**. (A) Effect of the speed of the solidification front on the thickness of the lamellae for alumina samples fabricated from powders with an average grain size of 0.3 µm (B to E). The scanning electron micrographs shown in the graph correspond to cross-sections parallel to the direction of movement of the ice front. Sample (E) was obtained with ultrafast freezing to gauge the thickness limit achievable by this technique. The approximate ice front velocity for this extreme case is in agreement with the extrapolation of the controlled freezing results. In addition, it is possible to control the materials mesostructure, e.g., in alumina (F), by patterning the surface of the cold fingers on which the ice crystals grow. Scale bars (B) 50 µm (C) 10 µm (D) 5 µm (E) 5 µm (F) 500 µm.

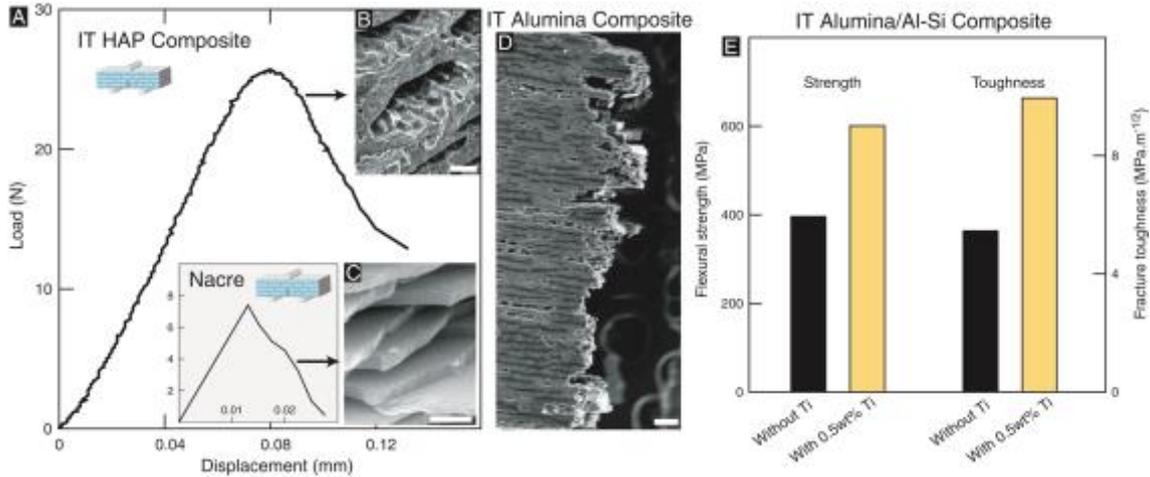

**Figure 3**. **Mechanical response of natural and synthetic IT composites.** The 3-points bending load-displacement data for IT HAP/epoxy composites (A) was qualitatively very similar to that of nacre (C) (17), with a gradually decreasing load after the elastic limit—characteristic of a stable crack propagation and active toughening—for cracks propagating in the direction perpendicular to the inorganic layers. Typical scanning electron micrographs of the IT composites (B) and nacre of abalone shell (C) reveal similar features on the fracture surface, with mode I cracks moving away from the notch and deflecting at the lamellae. Extensive crack deflection at the organic/inorganic interface results in tortuous crack paths and contributes to the toughening in both cases (as can be observed for the IT alumina/epoxy composite in D). The role of the interfacial chemistry, in the bonding between layers and the final mechanical properties of the material, is illustrated in the data shown in (E) for alumina/Al-Si composites (45/55 vol%), the addition of 0.5 wt% titanium to the aluminum alloy significantly increases the strength and toughness of the materials. Scale bars (B) 40 μm (C) 1 μm (D) 100 μm.

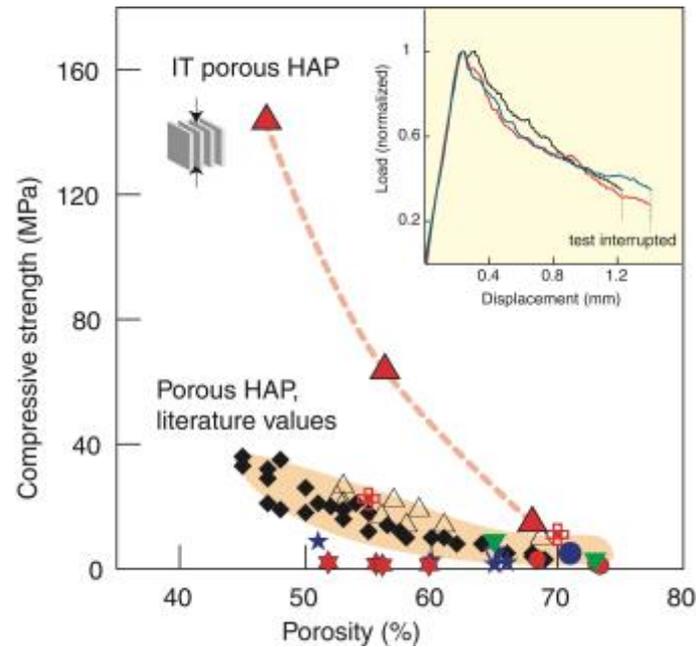

**Figure 4**. **Compressive strength of porous hydroxyapatite scaffolds**. Results from literature (22-30) vs. IT porous HAP scaffolds. The typical pore sizes of conventional porous HAP scaffolds are of the order of 100 μm to 800μm, in order to allow bone ingrowth. In the IT materials exhibiting the greatest strength, the pores are typically ~20 by ~200 microns wide and several millimetres long—previous studies have indicated that these dimensions are adequate for bone tissue engineering (20). For the IT porous materials, compression is applied in the direction parallel to the ceramic layers. Each style of points corresponds to a different source in the literature. The presence of inorganic bridges between the ceramic layers (a feature that parallels the microstructure of nacre) prevents Euler buckling of the ceramic layers and contributes to the high strength. The insert shows typical compression load-displacement curves for materials with 56% porosity (three different samples shown here). The samples fail gradually, and due to the large degree of control of hierarchical architecture, the mechanical behavior is very consistent from one sample to another.

**Supporting Online Material**
    www.sciencemag.org
    Materials and Methods
    Figs. S1, S2